\documentclass[a4paper,fleqn,usenatbib]{mnras}
\usepackage{comment}
\usepackage{dcolumn}

\usepackage[T1]{fontenc}
\usepackage{ae,aecompl}

\usepackage{graphicx}	
\usepackage{amsmath}	
\usepackage{amssymb}	
\usepackage{xspace}

\newcommand{\wav}{cm$^{-1}$\xspace}
\newcommand{\csixty}{C$_{60}$\xspace}

\title[Fullerenes in space]{Searching for stable fullerenes in space with computational chemistry}

\author[A. Candian et al.]{Alessandra Candian,$^{1}$\thanks{E-mail: candian@strw.leidenuniv.nl} Marina Gomes Rachid,$^{1,2}$
Heather MacIsaac,$^{3,4,5}$ \newauthor Viktor N. Staroverov,$^{6}$ Els Peeters$^{3,4,7}$
and Jan Cami$^{3,4,7}$ \\ \\
$^{1}$ Leiden Observatory, Leiden University, Niels Bohrweg 2, 2333CA, Leiden, The Netherlands \\
$^{2}$ Universidade do Vale do Para\'{i}ba, Laboratorio de Astroqu\'{i}mica e Astrobiologia,
   2911, Urbanova, S\~{a}o Jos\'{e} dos Campos, SP, Brazil \\
$^{3}$ Department of Physics and Astronomy, The University of Western Ontario, London, ON N6A 3K7, Canada \\
$^{4}$ Centre for Planetary Science and Exploration (CPSX), The University of Western Ontario, London, ON N6A 3K7, Canada \\
$^{5}$ Physics Department, St.~Francis Xavier University, 5005 Chapel
Square, Antigonish, NS B2G 2W5, Canada \\
$^6$ Department of Chemistry, The University of Western Ontario, London, ON N6A 5B7, Canada \\
$^{7}$ SETI Institute, 189 Bernardo Avenue, Suite 100, Mountain View, CA 94043, USA}

\date{Accepted XXX. Received YYY; in original form ZZZ}

\pubyear{2019}

\begin{document}
\label{firstpage}
\pagerange{\pageref{firstpage}--\pageref{lastpage}}
\maketitle

\large


\begin{abstract} 
We report a computational study of the stability and infrared (IR)
vibrational spectra of neutral and singly ionised fullerene cages
containing between 44 and 70 carbon atoms. The stability is
characterised in terms of the standard enthalpy of formation per CC
bond, the HOMO--LUMO gap, and the energy required to eliminate a C$_2$
fragment. We compare the simulated IR spectra of these fullerene
species to the observed emission spectra of several planetary nebulae
(Tc~1, SMP SMC 16, and SMP LMC 56) where strong \csixty emission has
been detected. Although we could not conclusively identify fullerenes
other than C$_{60}$ and C$_{70}$, our results point to the possible
presence of smaller (44, 50, and 56-atom) cages in those astronomical
objects. Observational confirmation of our prediction should become
possible when the James Webb Space Telescope comes online.
\end{abstract}




\section{Introduction}

A significant fraction ($\sim$10\%) of elemental carbon in the
universe is thought to exist in the form of large organic molecules
such as polycyclic aromatic hydrocarbons (PAHs) and fullerenes
\citep{2008ARA&A..46..289T}. These species are of central importance in
the physics and chemistry of interstellar environments and star-forming
regions of the Milky Way and other galaxies.

Fullerenes are a fairly recent addition to the inventory of known
interstellar species. Since the first detection of infrared (IR)
vibrational bands of C$_{60}$ and C$_{70}$ in the emission of the Tc 1
planetary nebula  (PN) \citep[]{Cami:C60-Science}, similar IR bands have been
found in many astronomical objects: various types of evolved stars
\citep{Garcia-Hernandez:PN, Gielen:C60p-AGB, ZhangKwok:proto-PNC60,
  Garcia-Hernandez:RcrB, Garcia-Hernandez:MC, 2012ApJ...760..107G,
  2012MNRAS.421L..92E, Jero:C60excitation}, reflection nebulae and
H{\sc ii} regions \citep{2010ApJ...722L..54S, 2012ApJ...760..107G,
  2012ApJ...753..168B, Peeters:2023, Pablo:PDRs}, young stellar
objects \citep{Roberts:C60}, and recently in the diffuse interstellar
medium \citep{2017A&A...605L...1B}. Laboratory measurements and
observational analyses have also shown that several diffuse
interstellar bands are due to C$_{60}^+$ \citep{Walker:C60+DIBs,
  Campbell:C60+DIBs, 2016ApJ...822...17C, 2016ApJ...831..130W,
  2017ApJ...843...56W, 2017ApJ...843L...2C, Lallement2018}. It has
thus become clear that C$_{60}$ and C$_{60}^+$ are widespread and abundant in space. 
Given that their unique spectral features are well characterized and understood, these species can now be used as proxies to study the much larger family of bulky aromatic species in space. .

A key question to be resolved in the context of circumstellar and
interstellar fullerenes is how they are formed. Most experimental
methods on Earth represent bottom-up formation routes starting from a
carbonaceous seed gas \citep[see, e.g.,][]{2009ApJ...696..706J}; closed
network growth through intermediate size cages eventually reaches the
most stable fullerene C$_{60}$ \citep{Dunk:CNG}. Such formation routes
require high densities and thus would play out over prohibitively long
timescales in astrophysical environments where densities are low
\citep[see, e.g.,][]{Elisabetta:arophatics}. An important clue to the
astrophysical formation routes came from the observational analysis of
\citet{2012PNAS..109..401B} who showed that the abundance of PAHs in
NGC~7023 decreases as one approaches the hot central star, while at
the same time the abundance of C$_{60}$ increases. This suggests that
fullerenes may be formed by ultraviolet photochemistry in a top-down
fashion starting from large PAHs with more than 60 carbons. The
chemical feasibility of such formation routes was confirmed
experimentally \citep{Zhen:formation_lab}. A top-down route may
explain the formation of C$_{60}$ in interstellar environments;
however, it is not clear whether it can also account for the formation
of fullerenes in evolved stars \citep{Cami+2018}.

Bottom-up routes toward \csixty  pass through smaller fullerene cages
that are intermediate building blocks;  top-down routes also involve
 smaller fullerene cages produced by
photofragmentation of \csixty. Interestingly, both bottom-up
\citep[e.g.][]{Zimmerman} and top-down \citep[e.g.][]{Rohlfing} routes
predict enhanced abundances of so-called ``magic-number" fullerenes --
smaller cages (with 44, 50 or 56 carbon atoms) that are more stable
than other fullerenes. If these intermediate fullerenes are
sufficiently stable, they may survive for long enough to be detectable
in astrophysical environments.

In this paper, we calculate and analyse several structural stability
indicators of various fullerenes which  may be  correlated with  their
abundance in space. We also calculate IR vibrational spectra for a
representative sample of intermediate fullerenes and compare our
results to astronomical observations of several \csixty-rich PNe.

\begin{table*}
\caption{Calculated properties of the C$_n$ and C$_n^{+}$ fullerene
  cages studied in this work.  All values were obtained at the
  B3LYP/6-31G* level of theory except for the HOMO--LUMO gap
  (PBE/6-31G*).}
\label{tab:cages}
\begin{tabular}{p{0.8cm}rccccD{.}{.}{2.2}cccD{.}{.}{4.1}D{.}{.}{2.2}} \hline
 & \multicolumn{6}{c}{Neutral C$_n$} & & \multicolumn{4}{c}{Cation C$_n^{+}$} \\ \cline{2-7} \cline{9-12}
  & & Point & $E_0$ & $\Delta_f H_0^\circ$ & $\Delta\epsilon_\text{HL}$ & \multicolumn{1}{c}{$\Delta E_0^\text{elim}$}
 && Point & $E_0$ & \multicolumn{1}{c}{$\Delta_f H_0^\circ$} & \multicolumn{1}{c}{$\Delta E_0^\text{elim}$} \\
 $n$ & Isomer$^{\ast}$ & group & ($E_h$) & (kcal/mol) 
 & (eV) & \multicolumn{1}{c}{(eV)}
 && group & ($E_h$) & \multicolumn{1}{c}{(kcal/mol)} & \multicolumn{1}{c}{(eV)} \\ \hline
 44 & 75 & $D_2$    & $-1675.934983$ & 823.3 & 0.76 & 8.51 && $C_2$ & $-1675.67508$0 &  986.3 & 8.51 \\
    & 89 & $D_2$    & $-1675.933854$ & 824.0 & 0.84 & 8.48 && $D_2$ & $-1675.672880$ &  987.7 & 8.45 \\
    & 72 & $D_{3h}$ & $-1675.922892$ & 830.8 & 1.23 & 8.18 && $C_1$ & $-1675.659056$ &  996.4 & 8.07 \\
    & 69 & $C_1$    & $-1675.905136$ & 842.0 & 0.57 & 7.70 && $C_1$ & $-1675.651777$ & 1001.0 & 7.87 \\
    & 87 & $C_2$    & $-1675.890008$ & 851.4 & 0.61 & 7.29 && $C_2$ & $-1675.640556$ & 1008.0 & 7.57 \\
    & 78 & $C_1$    & $-1675.887802$ & 852.9 & 0.43 & 7.23 && $C_1$ & $-1675.640151$ & 1008.3 & 7.56 \\
    & 88 & $C_1$    & $-1675.887805$ & 852.9 & 0.43 & 7.23 && $C_1$ & $-1675.639885$ & 1008.4 & 7.55 \\ [1.5ex]
 50 & 270 & $D_3$    & $-1904.622648$ & 840.7 & 1.33 & 9.57 && $C_2$    & $-1904.360527$ & 1014.3 & 9.21 \\
    & 271 & $D_{5h}$ & $-1904.614508$ & 845.8 & 0.41 & 9.35 && $C_{5h}$ & $-1904.372567$ & 997.6 & 9.54 \\
    & 266 & $C_{s}$  & $-1904.610068$ & 848.6 & 0.92 & 9.23 && $C_s$    & $-1904.357248$ & 1007.2 & 9.12 \\
    & 263 & $C_{2}$  & $-1904.606111$ & 851.1 & 1.13 & 9.13 && $C_2$    & $-1904.344217$ & 1015.4 & 8.77 \\
    & 264 & $C_{s}$  & $-1904.594589$ & 858.3 & 0.78 & 8.81 && $C_2$    & $-1904.342077$ & 1016.8 & 8.71 \\
    & 260 & $C_{2}$  & $-1904.587465$ & 862.8 & 0.54 & 8.62 && $C_s$    & $-1904.340925$ & 1017.5 & 8.68 \\
    & 262 & $C_{s}$  & $-1904.584073$ & 864.9 & 0.58 & 8.53 && $C_s$    & $-1904.334007$ & 1021.8 & 8.49 \\ [1.5ex]
 56 & 916 & $D_2$    & $-2133.263470$ & 887.6 & 0.68 & 8.36 && $C_s$    & $-2133.019324$ & 1040.8 & 8.31 \\
    & 864 & $C_s$    & $-2133.263101$ & 887.8 & 0.88 & 8.35 && $C_s$    & $-2133.017361$ & 1042.0 & 8.54 \\
    & 843 & $C_2$    & $-2133.257250$ & 891.5 & 0.70 & 8.19 && $C_2$    & $-2133.012619$ & 1045.0 & 8.13 \\
    & 913 & $C_{2v}$ & $-2133.252127$ & 894.7 & 0.50 & 8.05 && $C_{2v}$ & $-2133.011996$ & 1045.4 & 8.11 \\ [1.5ex]
 60 &     & $I_h$ & $-2285.799255$ & 850.7 & 1.67 & 11.18 && $D_{5d}$ & $-2285.538791$ & 1014.1 & 10.47 \\ [1.5ex]
 62 & 1h$^{\dagger}$  & $C_s$ & $-2361.916462$ & 926.8 & 0.50 & 5.52 && $C_s$    & $-2361.681847$ & 1074.0 & 6.22 \\
    & 1s$^{\ddagger}$ & $C_{2v}$ & $-2361.910879$ & 930.3 & 0.84 & 5.36 && $C_{2v}$ & $-2362.670169$ & 1081.3 & 5.90 \\
    & 2378 & $C_2$            & $-2361.895289$ & 940.1 & 0.42 & 4.93 && $C_1$    & $-2361.655576$ & 1090.5 & 5.50 \\
    & 2377 & $C_1$            & $-2361.894581$ & 940.5 & 0.34 & 4.92 && $C_2$    & $-2361.653374$ & 1091.5 & 5.44 \\ [1.5ex]
 64 & 3451 & $D_2$ &$-2438.155785$ & 926.3 & 1.23 & 8.84 && $D_2$ & $-2437.907149$ & 1082.3 & 9.17 \\
    & 3452 & $C_s$ &$-2438.145459$ & 932.7 & 1.06 & 8.56 && $C_s$ & $-2437.894842$ & 1090.0 & 8.83 \\
    & 3457 & $C_2$ &$-2438.135884$ & 938.8 & 0.86 & 8.30 && $C_2$ & $-2437.885889$ & 1095.6 & 8.59 \\ [1.5ex]
 66 & 4466 & $C_s$    & $-2514.356701$ & 949.8 & 0.95 & 7.79 && $C_{2v}$ & $-2514.113257$ & 1102.6 & 7.93 \\
    & 4348 & $C_{2v}$ & $-2514.350102$ & 954.0 & 0.30 & 7.61 && $C_s$    & $-2514.106888$ & 1106.6 & 7.76 \\
    & 4169 & $C_2$    & $-2514.342759$ & 958.6 & 0.48 & 7.41 && $C_2$    & $-2514.099333$ & 1111.3 & 7.55 \\ [1.5ex]
 68 & 6290 & $C_2$ & $-2590.574909$ & 962.6 & 1.37 & 8.26 && $C_2$ & $-2590.323827$ & 1120.1 & 8.06 \\
    & 6328 & $C_2$ & $-2590.572785$ & 963.9 & 0.93 & 8.21 && $C_2$ & $-2590.324464$ & 1119.7 & 8.07 \\
    & 6270 & $C_1$ & $-2590.557981$ & 973.2 & 1.05 & 7.80 && $C_1$ & $-2590.313048$ & 1126.9 & 7.76 \\
    & 6198 & $C_1$ & $-2590.555998$ & 974.4 & 0.41 & 7.75 && $C_1$ & $-2590.317355$ & 1124.2 & 7.88 \\
    & 6148 & $C_1$ & $-2590.554346$ & 975.5 & 0.36 & 7.71 && $C_1$ & $-2590.321569$ & 1121.6 & 7.99 \\
    & 6146 & $C_2$ & $-2590.554036$ & 975.7 & 0.07 & 7.69 && $C_2$ & $-2590.315973$ & 1125.1 & 7.84 \\
    & 6195 & $C_2$ & $-2590.553849$ & 975.8 & 0.42 & 7.70 && $C_2$ & $-2590.328153$ & 1117.4 & 8.17 \\
    & 6094 & $C_s$ & $-2590.553605$ & 975.9 & 0.04 & 7.68 && $C_s$ & $-2590.319309$ & 1123.0 & 7.93 \\ [1.5ex]
 70 &      & $D_{5h}$ & $-2666.865607$ & 929.8 & 1.70 & 10.24 && $C_{2v}$ & $-2667.046354$ & 1089.6 & 10.02 \\ \hline
\multicolumn{12}{l}{$^{\ast}$Isomer identification number according to \citealt{FowlerMan}.} \\
\multicolumn{12}{l}{$^{\dagger}$Structure with a heptagon.} \\
\multicolumn{12}{l}{$^{\ddagger}$Structure with a square.} \\
\end{tabular}
\end{table*}

\section{Methodology}

\subsection{Definitions}

The stability of fullerene-type molecules is determined by several
factors. By Euler's polyhedron formula, a fullerene consisting only of
pentagons and hexagons (classical fullerenes) must contain exactly 12
pentagons to form a closed cage. The arrangement of the pentagons
affects the stability of the cage, the most stable isomer being the
one where the pentagons are isolated \citep{Kroto.1987.N.329.529}. The
buckminsterfullerene C$_{60}$ and its larger cousin C$_{70}$ are the
smallest cages that comply with this rule \citep{FowlerMan}. For cages
where isolated pentagons are impossible, the more stable isomers are
the ones with the smallest number of adjacent pentagons
\citep{Albertazzi.1999.PCCP.1.2913}.  Although the isolated pentagon rule is likely the dominant stabilisation factor  for neutral and cationic cages, there are indications that it does not always work \citep{Fowler.1992.N.355.428, Wang2015}.

Here, we are interested in the relative stability of smaller C$_n$
cages with $n=44$, 50, 56 as well as cages that arise as intermediates
in reversible transformations connecting C$_{60}$ and C$_{70}$, i.e.,
$n=62$, 64, 66, and 68. Guided by the  isolated pentagon rule, we
selected for each $n$ the isomers containing the lowest number of
adjacent pentagons in both neutral and cationic form. In the case of
C$_{62}$, non-classical fullerenes containing a heptagon and square
were also considered, since they are expected to be more stable than
the classical cages \citep{Ayuela1996, Qian2000, Sanchez2005}.

To quantify the stability of these C$_n$ fullerene molecules, we
employ three metrics.  The first one is $\Delta_f H_0^\circ$, the standard
enthalpy of formation of C$_n$ at $T=0$ K divided by the number of CC
bonds. $\Delta_f H_0^\circ$ is defined as the change of enthalpy for
the reaction in which 1 mol of C$_n$ is formed from $n$ mol of
graphite at 1~bar.  Following \citet{Alcami2007}, we compute this
quantity using the relation
\begin{equation}
 \label{form:DeltaH_f}
  \Delta_f H_0^\circ(\mbox{C$_n$})
  = E_0(\mbox{C$_n$}) - n E_0(\mbox{C}) + n \Delta_f H_0^\circ(\mbox{C}),
\end{equation}
where $E_0(\mbox{C$_n$})$ is the sum of the total electronic
and zero-point energies (ZPE) of C$_n$, $E_0(\mbox{C})$ is the
ground-state electronic energy of a gas-phase C atom, and $\Delta_f
H_0^\circ(\mbox{C})=171.29$ kcal/mol is the standard enthalpy of
formation of gas-phase C atoms at $T=0$ K. The values of
$E_0(\mbox{C$_n$})$ and $E_0(\mbox{C})$
were calculated with quantum chemistry techniques (see  below).

Stability of fullerenes can also be described from a kinetic point of
view as a measure of ``resistance" to becoming an activated complex
that can undergo chemical reactions such as fragmentation.  A useful
indicator of kinetic stability is the gap between the highest occupied
molecular orbital (HOMO) and lowest unoccupied molecular orbital
(LUMO) energy levels,
\begin{equation}
  \label{eq:HOMO-LUMO-gap}
 \Delta \epsilon_\text{HL} = \epsilon_\text{LUMO} - \epsilon_\text{HOMO}.
\end{equation}

\noindent A large HOMO--LUMO gap correlates with a high kinetic stability (lower
chemical reactivity) of C$_n$ molecules
\citep{Manolopoulos.1991.CPL.181.105}.

Finally, molecular stability can be interpreted as the enthalpy change
associated with a particular fragmentation reaction. For example, it
is well established that the dominant fragmentation channel of C$_{n}$ cages
is a sequential elimination of C$_2$ units. The enthalpy change for this process is
\begin{equation}\label{form:e_diss}
 \Delta E_0^\text{elim}
  = E_0(\mbox{C$_{n-2}$}) + E_0(\mbox{C$_2$}) - E_0(\mbox{C$_{n}$}),
\end{equation}
where the $E_0$ values are the ZPE-corrected total electronic energies
of the indicated gas-phase species.

\subsection{Computational details}

Cartesian coordinates of all C$_n$ cages were generated using the
\texttt{Fullerene} software (version~4.5) \citep{Fullerene}, applying
symmetry constraints appropriate to the corresponding point group. We
optimised cage geometries and calculated their vibrational spectra
with the \texttt{Gaussian 09} program \citep{g09} using the B3LYP
functional \citep{Stephens.1994.JPC.98.11623}, the \mbox{6-31G*} basis
set, and a frequency scale factor \mbox{$f_\text{scale}=0.978$}. We
convolved the spectra with a Lorentzian profile function having a full
width at half maximum (FWHM) of 10~\wav, unless noted otherwise. We
used \texttt{Gabedit} \citep[version 2.5.0;][]{gabedit} to visualize
the cage structures and their vibrational modes. We chose the B3LYP
functional because it had been successfully used in the past to obtain
the IR spectra of other fullerenes \citep{Adjizian2016} and conjugated
hydrocarbons \citep{Langhoff1996} with various basis sets (4-31G,
6-31G, 6-31G*, 6-31G**, 6-311G). The B3LYP/6-31G* combination with the
above frequency scaling factor provides the best agreement with
experiment \citep{gas_phaseC60:Frum1991,IR_matrix:Kern2013,
  C70:Nemes1994}, similar to that of the B3LYP/6-311G method used by
\citet{Adjizian2016}. The C$_2$ elimination energy
 was also calculated at the B3LYP/6-31G* level. For each cage
C$_n$, we used the most stable isomer of C$_{n-2}$ as the
fragmentation product. The B3LYP/6-31G* method predicts the C$_2$
molecule to have a triplet ground state \citep{Diaz-Tendero2003},
which is the state we used here for computing $E_0(\mbox{C}_2)$ in
Eq.~\eqref{form:e_diss}.

Because the B3LYP functional performs poorly for HOMO--LUMO gaps of
C$_{60}$ \citep{Kremer1993}, we followed \citet{Beu2005} and
calculated orbital energies using the PBE functional \citep{PBE} with
the same 6-31G* basis set, at the PBE/6-31G* geometries.

\section{Results}

\subsection{Thermochemical stability}

\begin{figure}
\includegraphics[width=\columnwidth]{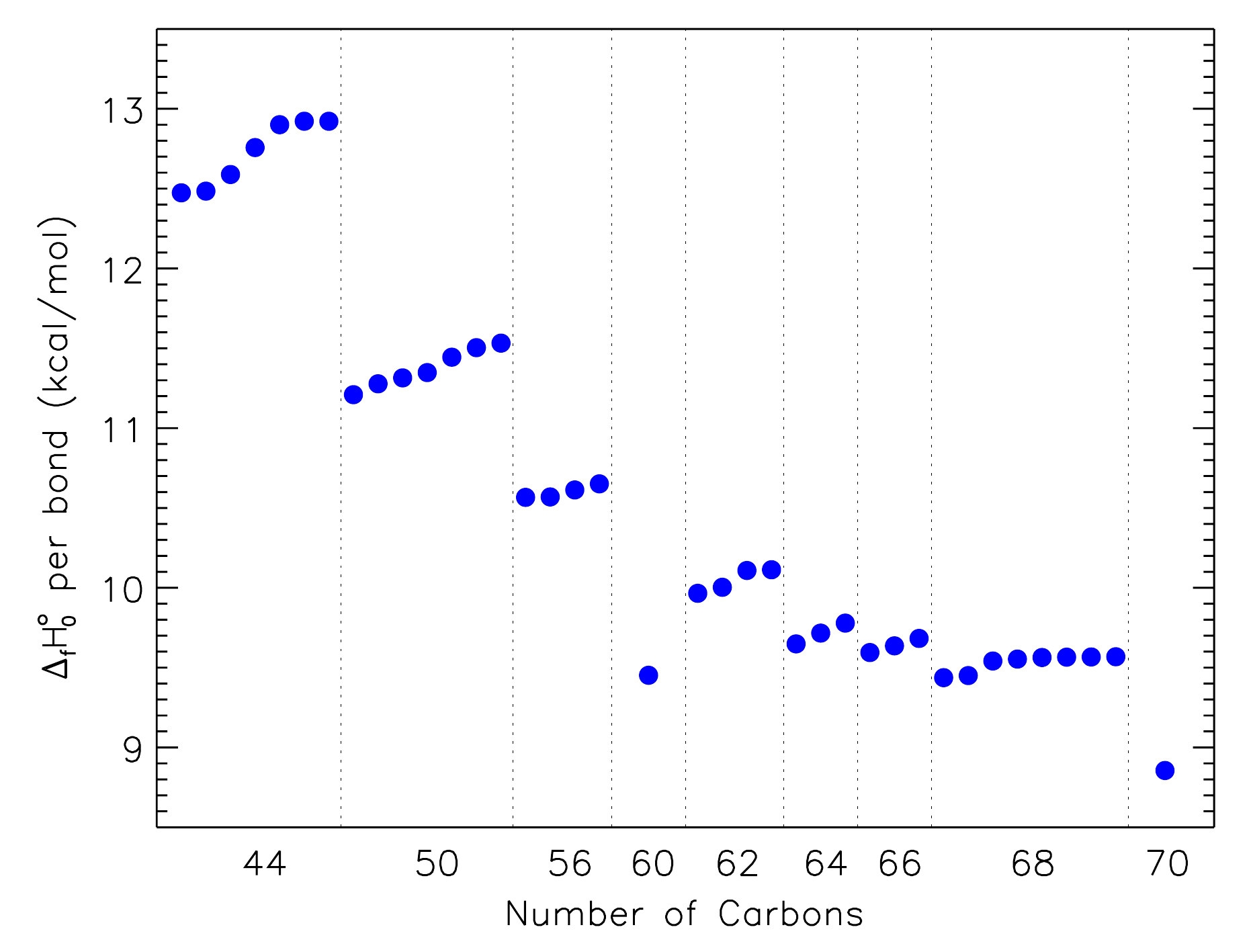}
\caption{Standard enthalpies of formation per CC bond for neutral C$_n$ cages, calculated with the
  B3LYP/6-31G* method. For each $n$, the isomers are ordered by their
  increasing total energy (as in Table~\ref{tab:cages}). Dashed
  vertical lines separate the species with a different number of
  carbon atoms.}
\label{fig:deltaH}
\end{figure}

\begin{figure}
\includegraphics[width=\columnwidth]{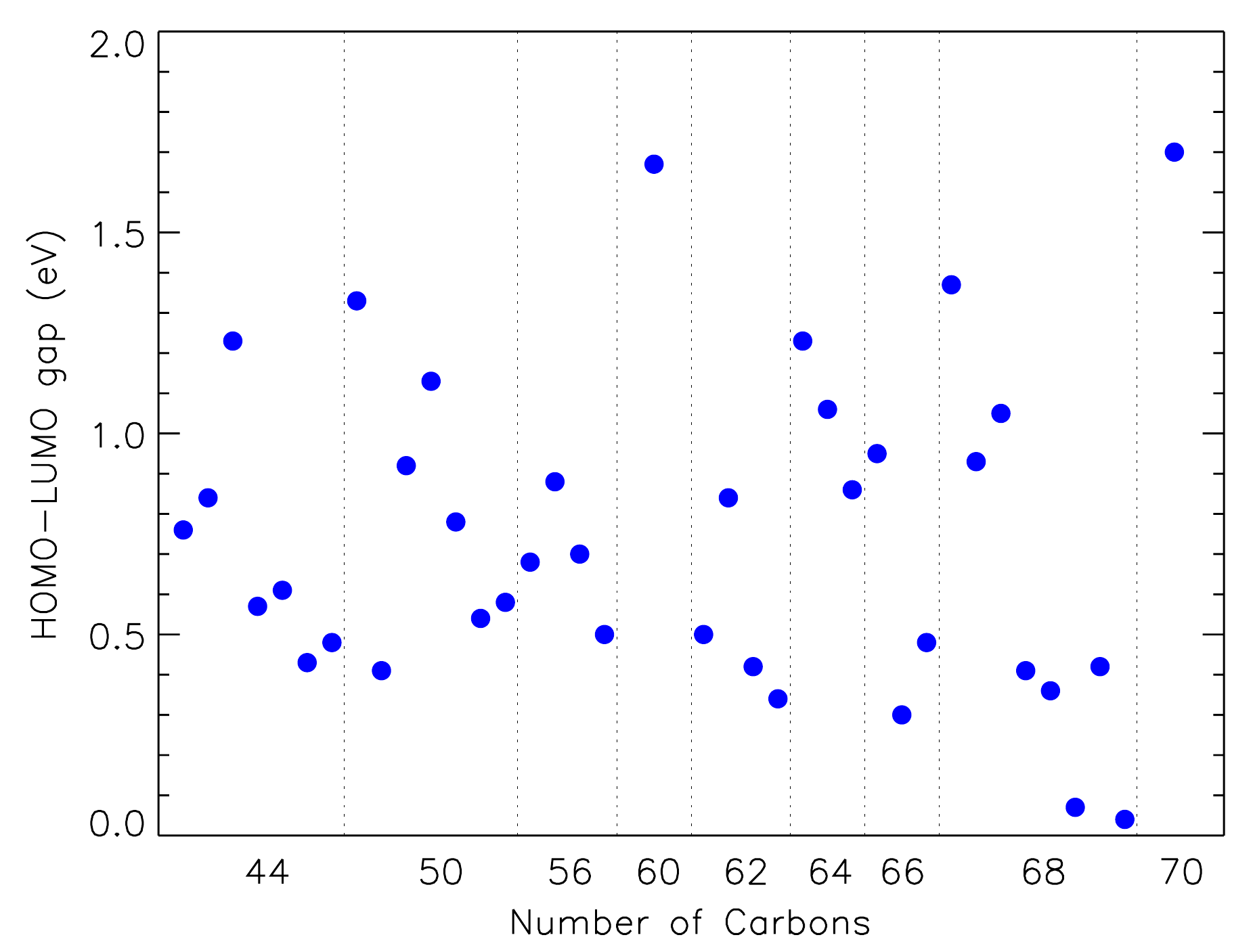}
\caption{The HOMO--LUMO gaps of neutral C$_n$ cages, calculated with the PBE/6-31G* method. The isomers are
  ordered and separated as in Fig.~\ref{fig:deltaH}. }
\label{fig:HL_gap}
\end{figure}

\begin{figure}
\includegraphics[width=\columnwidth]{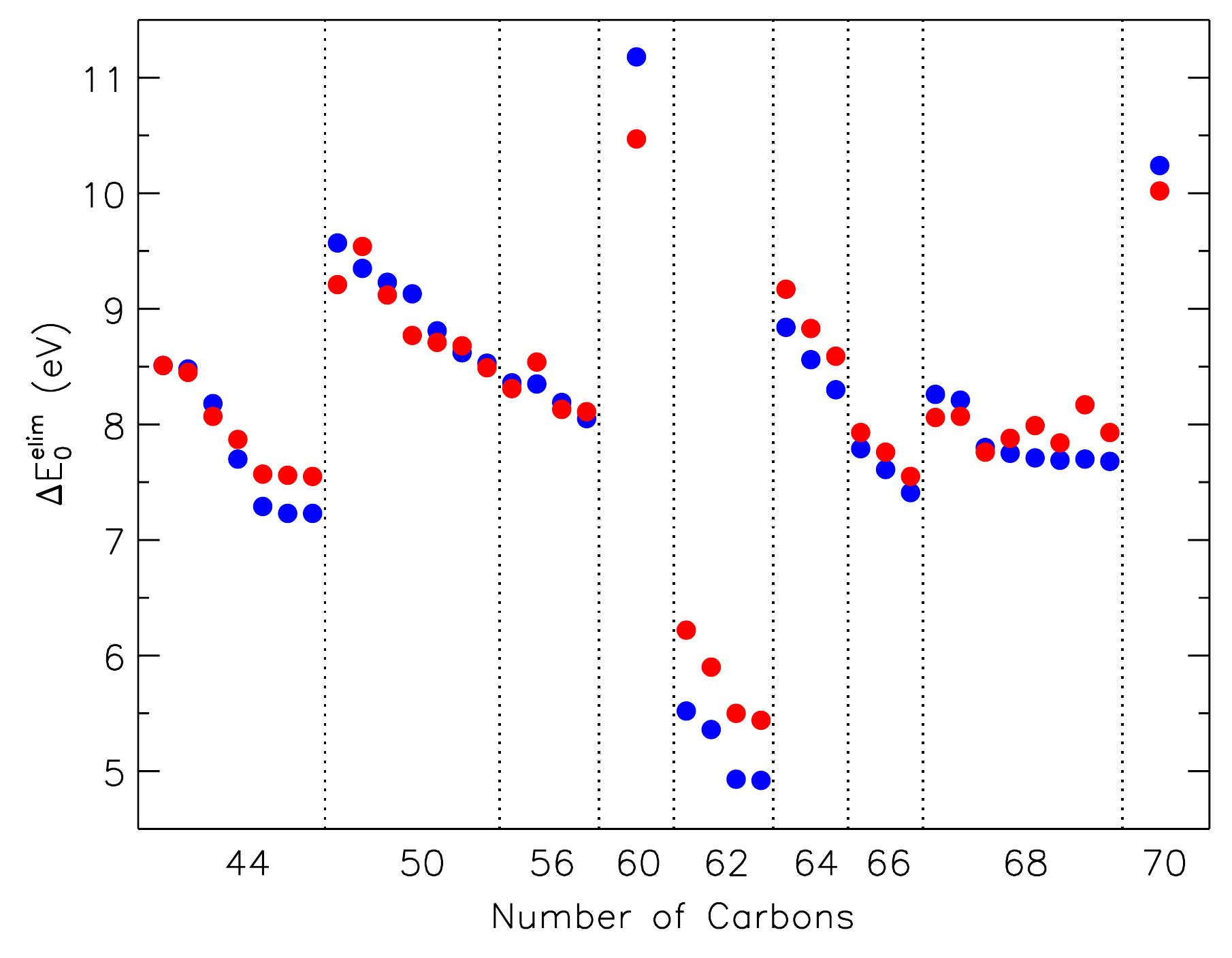}
\caption{C$_2$ elimination energies of various C$_n$ (blue circles)
and C$_n^{+}$ cages (red circles) calculated by the B3LYP/6-31G* method using
Eq.~\eqref{form:e_diss}. For each $n$, the isomers are ordered by their
increasing total energy. The order of stability for the C$_n^{+}$ cages
is the same as for corresponding neutral fullerenes with the exception
of $n=66$ and 68 (see Table~\ref{tab:cages}).
Dashed vertical lines are drawn to guide the eye.}
\label{fig:C2diss}
\end{figure}

We summarise the results of all thermochemical calculations of this
work in Table~\ref{tab:cages};
Figs.~\ref{fig:deltaH}--\ref{fig:C2diss} provide a graphical
representation of the same results.

Figure~\ref{fig:deltaH} shows the calculated enthalpy of formation per
CC bond (assuming $3n/2$ bonds in C$_n$) of the neutral C$_n$ cages 
considered. The results are in excellent agreement with a previous
study \citep{Alcami2007}. As expected, C$_{60}$ and C$_{70}$ are the
most stable structures according to this metric, but we note that
several isomers of C$_{68}$  are almost as stable as C$_{60}$. The general
decrease in $\Delta_f H_0^\circ$ (per bond) values observed from
C$_{40}$ to C$_{70}$ correlates with the decreasing geometric
distortion of the trigonal planar geometry preferred by each
$sp^2$-hybridised carbon. Variations of the $\Delta_f H_0^\circ$
values among isomers are small, especially for larger cages.

Figure~\ref{fig:HL_gap} shows the calculated  HOMO--LUMO gaps for
neutral C$_n$ cages. C$_{60}$ and C$_{70}$ have the largest gaps among
the structures examined: 1.67~eV and 1.71~eV, respectively. The
smallest gaps are close to zero (0.04 and 0.07 eV for isomers 6094
and 6146 of C$_{68}$). Note that, among the isomers with a given $n$,
the structures with the largest HOMO--LUMO gaps are not
always the most thermodynamically stable ones. Overall, the scatter of the
$\Delta\epsilon_\text{HL}$ values is considerably greater than that
of the standard enthalpies of formation per bond, which suggests
that kinetic and thermodynamic stabilities of C$_n$ cages are not
strongly correlated.

Figure~\ref{fig:C2diss} shows the C$_2$ elimination energies of C$_n$
cages in their neutral and cationic forms. In line with their
exceptional stability, neutral C$_{60}$ and C$_{70}$ require the
highest energy ($\sim$11 eV) to remove a C$_2$ unit, followed closely
by their cations. These results are in agreement with previous
calculations and experimental data \citep[and references
  therein]{Diaz-Tendero2003,Diaz-Tendero2006}. The next most stable
cages according to this metric are C$_{50}$ and C$_{64}$, while the
C$_{62}$ isomers are the least stable. This ordering is preserved for
the respective cations. Generally, C$_2$ elimination energies of the
C$_n^{+}$ cations are close to the $\Delta E_0^\text{elim}$ values of
the corresponding neutral C$_n$ structures.

\subsection{Infrared spectra}
\begin{figure*}
\includegraphics[scale=0.7]{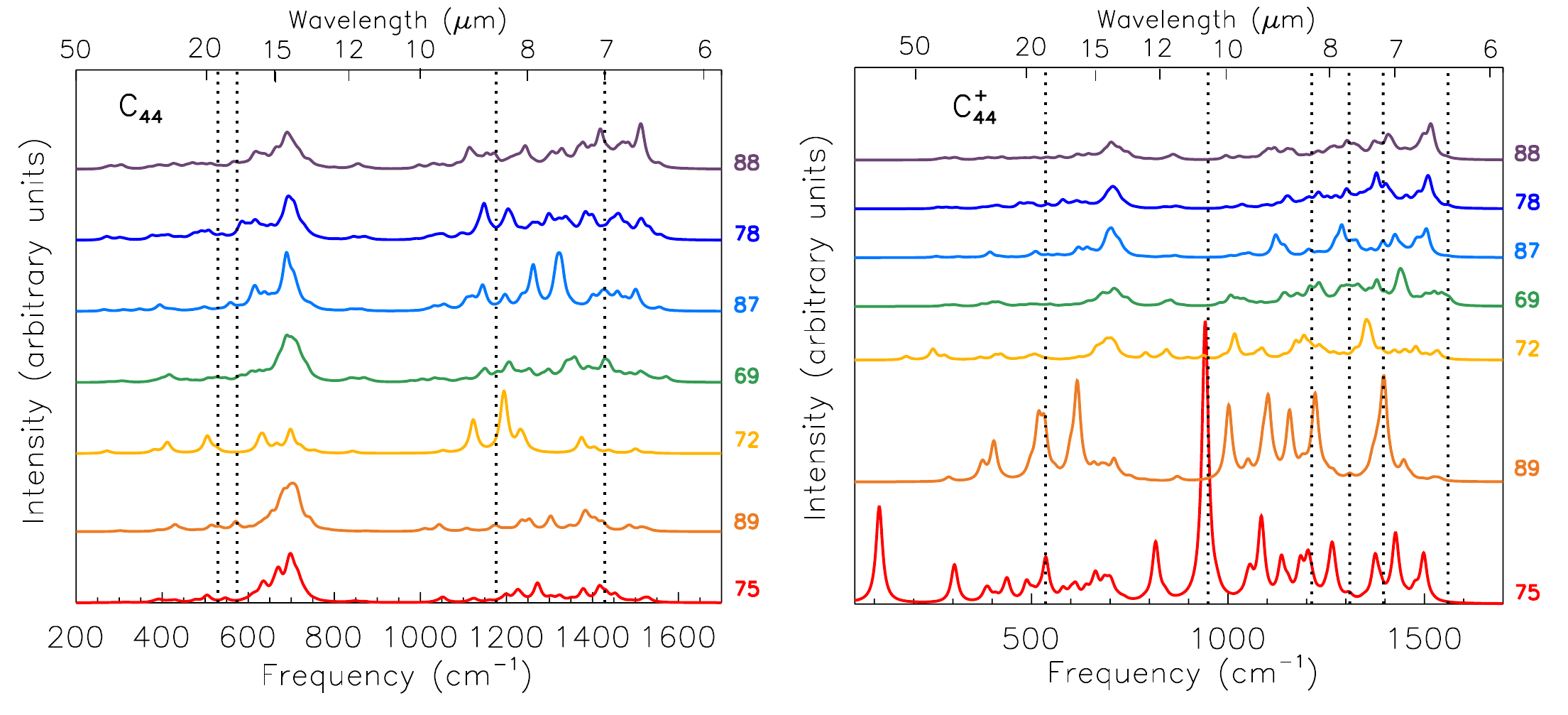}
\caption{Simulated IR absorption spectra of  various isomers of
  neutral C$_{44}$ (left panel) and of their cationic forms (right
  panel). The spectra are ordered from top to bottom as in Table~\ref{tab:cages}. The dotted lines show the
  positions of the IR-active modes of C$_{60}$ at 7.0, 8.5, 17.4, and
  18.9~$\mu$m (left) and C$_{60}^+$ at 6.4, 7.1, 7.5, 8.2, 10.4, and
  18.9~$\mu$m (right).}
\label{fig:C44}
\end{figure*}

\begin{figure*}
\includegraphics[scale=0.7]{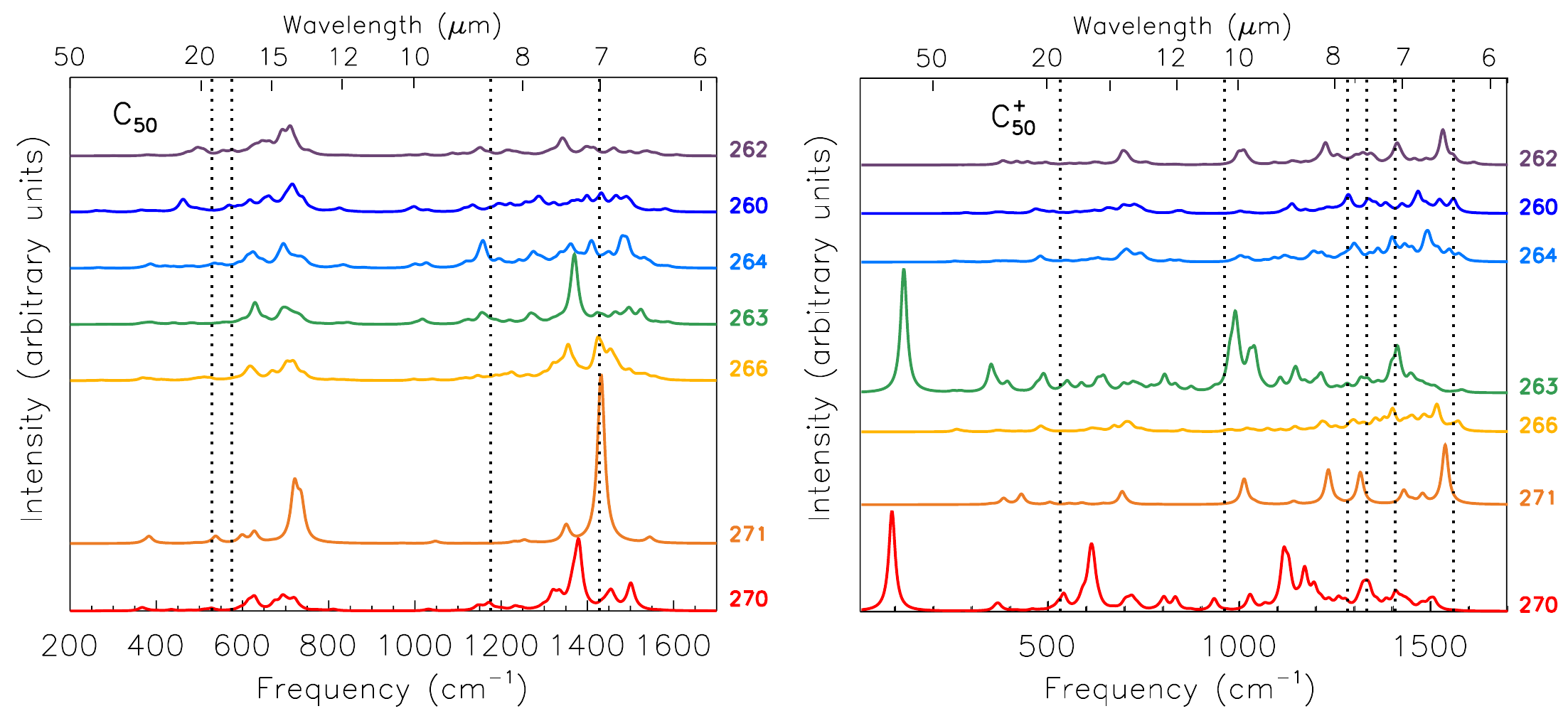}
\caption{Same as in Fig.~\ref{fig:C44} but for C$_{50}$ (left panel) and of
their cationic forms (right panel).}
\label{fig:C50}
\end{figure*}

\begin{figure*}
\includegraphics[scale=0.7]{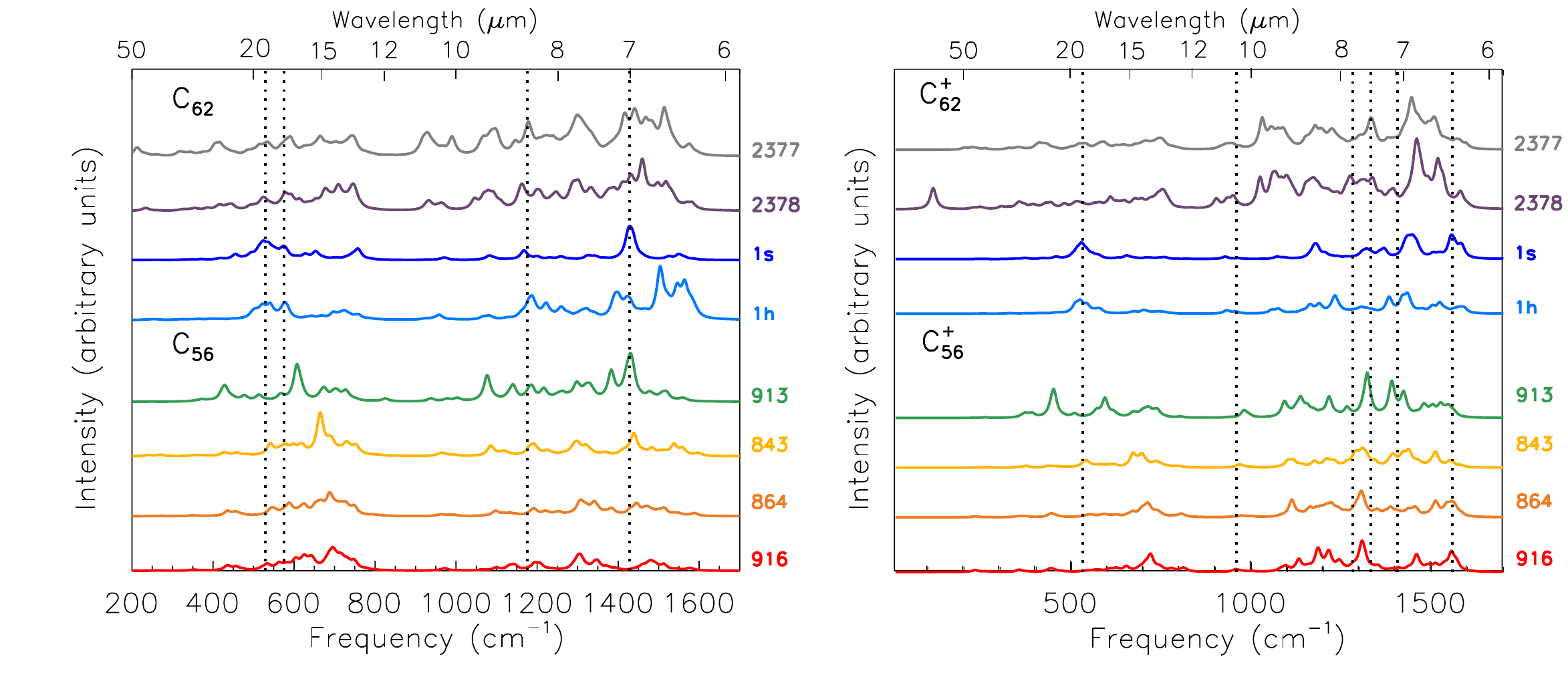}
\caption{Same as in Fig.~\ref{fig:C44} but for C$_{56}$ and C$_{62}$ (left panel)
and of their cationic forms (right panel).}
\label{fig:C56C62}
\end{figure*}

\begin{figure*}
\includegraphics[scale=0.7]{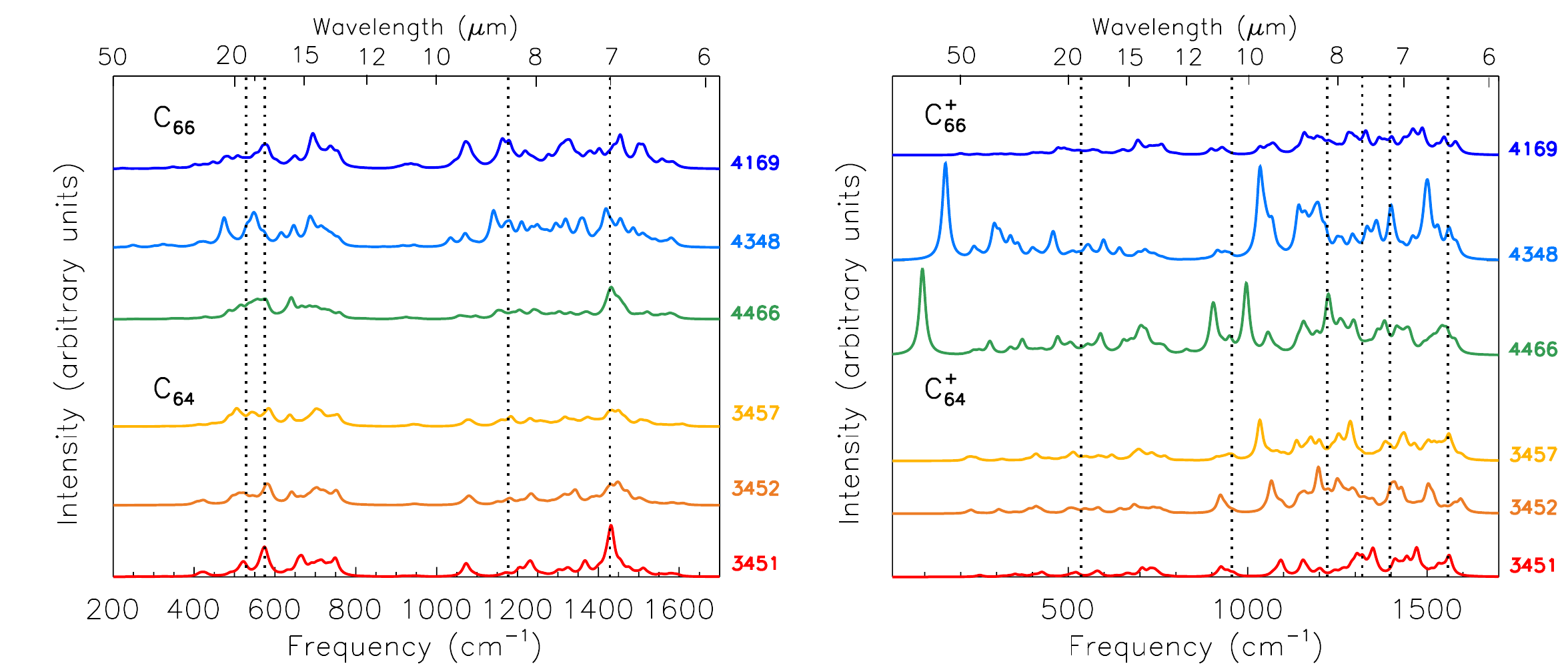}
\caption{Same as in Fig.~\ref{fig:C44} but for C$_{64}$ and C$_{66}$ (left panel)
and of their cationic forms (right panel).} 
\label{fig:C66C64}
\end{figure*}

\begin{figure*}
\includegraphics[scale=0.7]{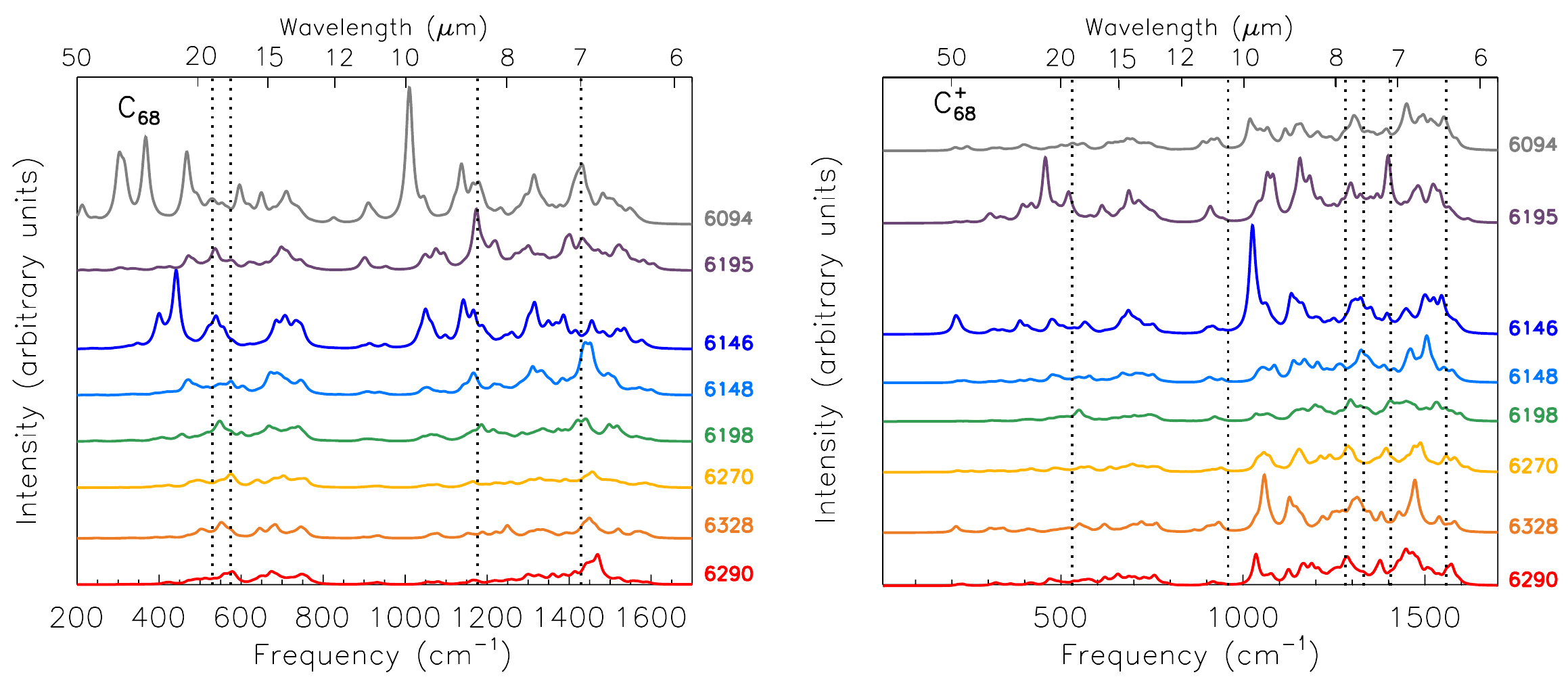}
\caption{Same as in Fig.~\ref{fig:C44} but for C$_{68}$ (left panel)
and of their cationic forms (right panel).} 
\label{fig:C68}
\end{figure*}

\begin{figure*}
\includegraphics[scale=0.7]{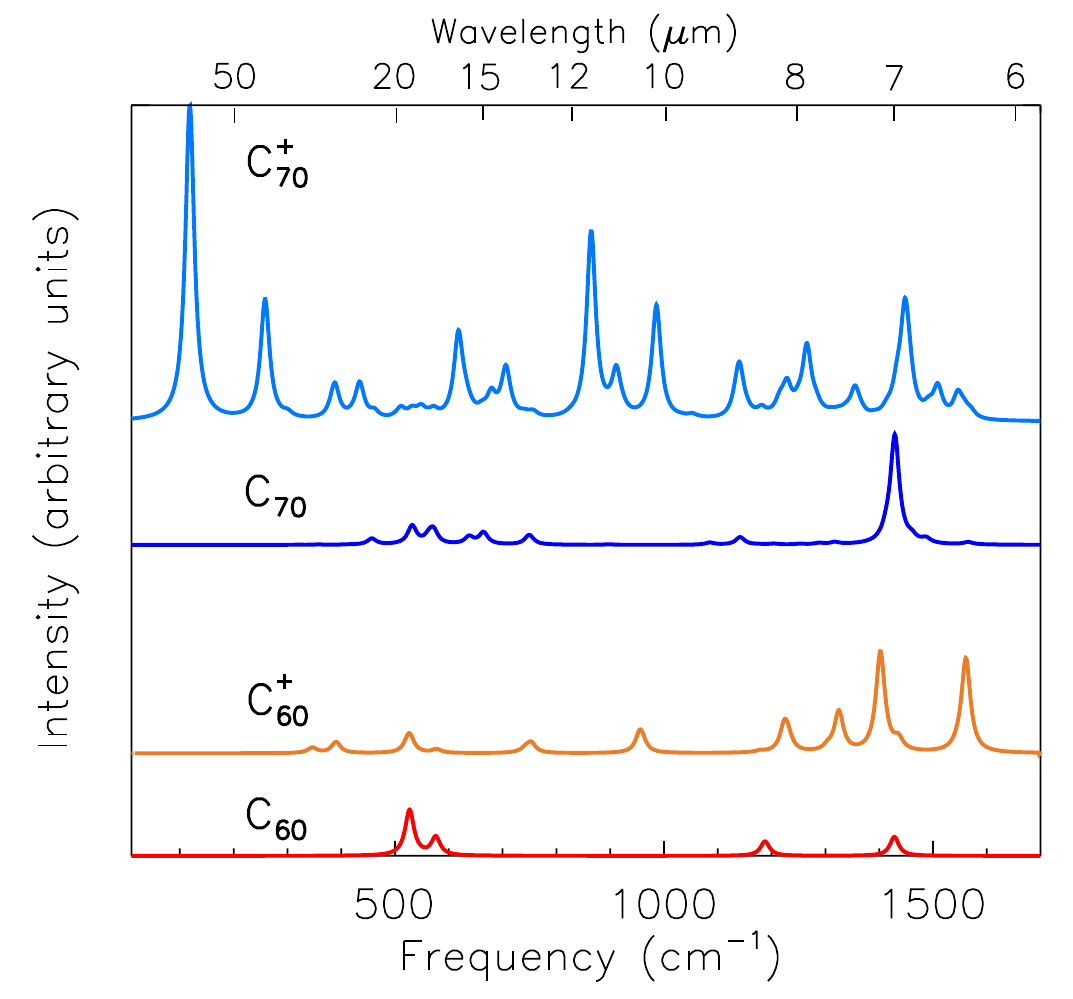}
\caption{IR spectra of C$_{60}$  and C$_{70}$ in neutral and cationic forms calculated  with the B3LYP/6-31G* method ($f_\text{scale}=0.978$),}
\label{fig:C60C70}
\end{figure*}

The full collection of the simulated spectra of the fullerenes of
Table~\ref{tab:cages} is displayed in
Figs.~\ref{fig:C44}--\ref{fig:C68}. Machine-readable tables with the
calculated spectra are available as Supplementary Material. At first
glance, IR spectra of all neutral cages display many
similarities. Most of the IR-active modes cluster in two spectral
regions: 1540--1000~\wav (6.5--10$~\mu$m) for CC stretching modes and
800--550~\wav (12.5--18.2~$\mu$m) for carbon-skeleton breathing modes.
This is consistent with the calculated spectra of smaller cages
($n=20$, 24, 26, 28, and 30) reported by \citet{Adjizian2016}. The
extent of the first spectral region is almost constant within our set
of spectra, while the second is subject to variations. This is a
consequence of the fact that the breathing modes are much more specific
to each species; therefore, this wavelength range can be considered as
a ``fingerprint region'' for individual species.

In several cases, IR active modes show up around 1000--900~\wav
(10.0--11.1~$\mu$m), due to combinations of CC stretching and
cage-breathing modes (see, for example, C$_{50}$ and C$_{68}$ in
Figs.~\ref{fig:C50} and \ref{fig:C68}). No strong or moderately strong
IR-active modes are found below 200~\wav (50~$\mu$m) and above
1600~\wav (6.25~$\mu$m). The detailed appearance of the IR spectrum
depends on the specific molecular structure and is influenced by
symmetry: the lower the symmetry, the larger the number of
low-intensity IR modes. This effect is exemplified by C$_{50}$ isomers
271 ($D_{5h}$) and 260 ($C_2$) in the left panel of Fig.~\ref{fig:C50}
-- the  isomer 271 has a much higher symmetry than the isomer  260 and
consequently exhibits only a few strong bands, whereas the isomer 260
shows a large number of weaker bands.

Neutral cages generally have the strongest IR transitions in the
665--714~\wav region (15.0--14.0~$\mu$m). Even when the strongest
transition is not in this range, it always contains a group of
superposed medium-intensity transitions, creating broad peaks above
what appears as a low-intensity plateau. Some notable exceptions are
the highly symmetric C$_{50}$ isomer 270 ($D_{5h}$) and C$_{70}$
($D_{5h}$), all showing their strongest modes around 7~$\mu$m
(1430~\wav). Other low-symmetry neutral cages may show strong modes
between 1540 and 1000~\wav, but then the integrated intensity in this
region is comparable to that of the breathing-mode region.

Few small cages ($n<60$), but almost all of the larger ones, show strong
modes close to the characteristic frequencies of C$_{60}$: 7.0, 8.5,
17.4, and 18.9~$\mu$m. For isomer 271 of C$_{50}$ and for C$_{70}$,
the band at 7.0~$\mu$m is the strongest in their spectra.

The spectra of the ionised cages resemble those of their neutral
counterparts, but generally show enhanced CC stretching modes
(1540--1000~\wav or 6.6--10~$\mu$m). Comparing the calculated absolute
intensities for neutral and ionised fullerene, one sees that this effect
is due to a decrease of the intensity of the skeleton-breathing modes at
longer wavelength. Moreover, there is a tendency for the modes to move
to shorter wavelength \citep{Adjizian2016}. The effect of ionisation
in fullerene thus mirrors what is seen in other large conjugated system
such as PAHs \citep[e.g.]{Bausch2008}.

Highly symmetric neutral C$_n$ cages have degenerate but completely
filled HOMOs, so they form nondegenerate ground states. After the
removal of an electron, the HOMO-level degeneracy comes into play and
the ionised cages spontaneously lower their symmetry due to the
Jahn--Teller distortion \citep{JahnTeller}. C$_{60}$ is particularly
notable in that regard \citep{Berne13,IR_matrix:Kern2013}. The lower
symmetry increases the complexity of the cation spectrum, as seen for
instance in isomer 271 of C$_{50}^+$ (Fig.~\ref{fig:C50}). In a few
cases (including isomer 75 of C$_{44}^+$, C$_{66}^+$, and C$_{70}^+$),
the IR spectrum predicted by B3LYP/6-31G* shows anomalously high IR
intensities, often for very low energy modes, as well as a complete
mismatch with the spectra of their neutral counterparts (see
Section~\ref{sec:challenges} for further discussion of this aspect).

\begin{figure}
\centering
\includegraphics[width=\columnwidth]{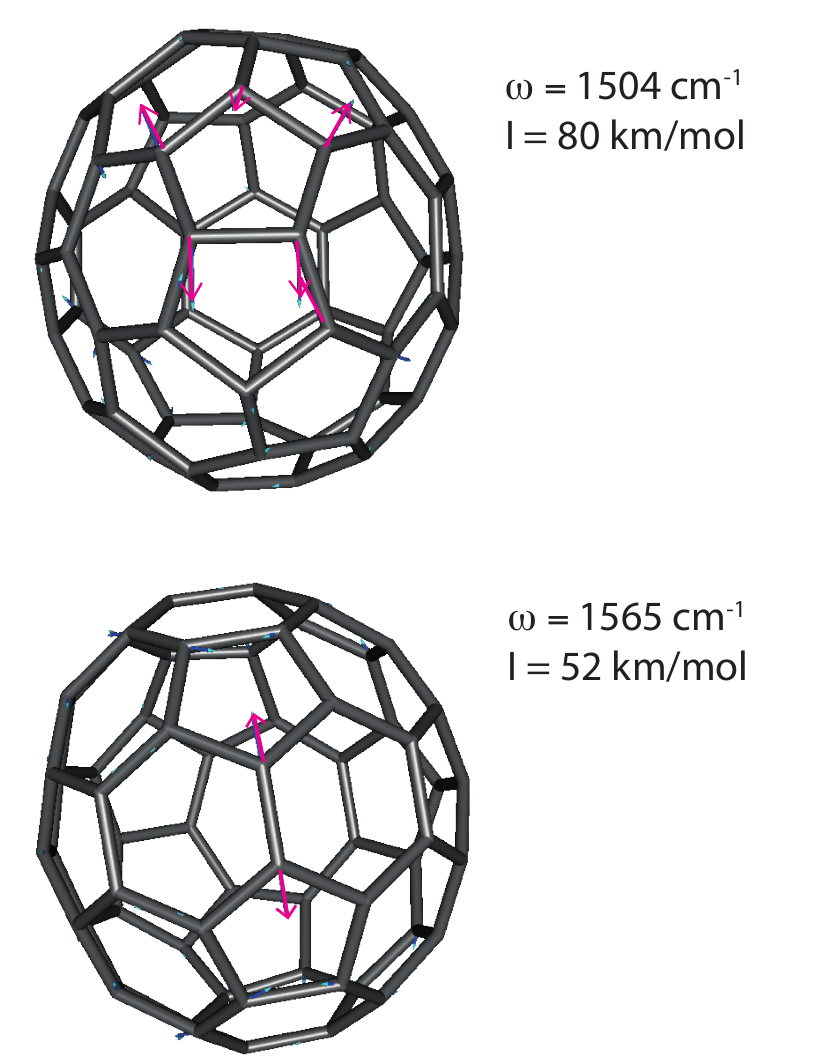}
\caption{Calculated normal vibrational modes associated with two intense IR peaks for isomer 1h of C$_{62}$. In the top image the 7-membered ring
is on the right-hand side of the pentagons involved in the mode; in
the bottom image the 7-membered ring can be seen on the back CC bond
involved in the mode.}
\label{fig:C62_motion}
\end{figure}

Finally, in our sample we have two non-classical isomers, a C$_{62}$
isomer containing a square 4-membered ring (1s) and a C$_{62}$ isomer
containing a 7-membered ring (1h). For isomer 1s, the spectrum has
only a few modes with very low intensities involving the square motif
and this does not change much upon ionisation. On the other hand,
isomer 1h has medium-strong peaks at 1179, 1222, 1387, and 1391 \wav
(8.52, 8.18, 7.21, and 7.19~$\mu$m) originating mostly from the
stretching of the 7-membered ring. Also, C$_{62}$-1h is the only
neutral cage in our sample showing strong signals below 1500 \wav, at
1504 \wav (6.65~$\mu$m), 1546 \wav (6.47~$\mu$m) and 1565 \wav
(6.39~$\mu$m) (Fig.~\ref{fig:C56C62}, left panel). While the first
mode is due to a CC stretch in the two pentagons sharing bonds with the
heptagons, the last two modes originate from the stretch of the CC bond
connecting two hexagons opposite of the 7-membered ring
(Fig.~\ref{fig:C62_motion}). Upon ionisation, the modes involving the
heptagon move slightly (10--20 \wav) to higher frequencies and
preserve their intensities. The modes lower than 1500 \wav
(6.67~$\mu$m) have decreased intensities by a factor of 2--3.

\subsection{Challenges of computing the IR spectrum of C$_{70}^+$  (and other ions)}
\label{sec:challenges}

\begin{figure*}
\centering
\includegraphics[scale=0.72]{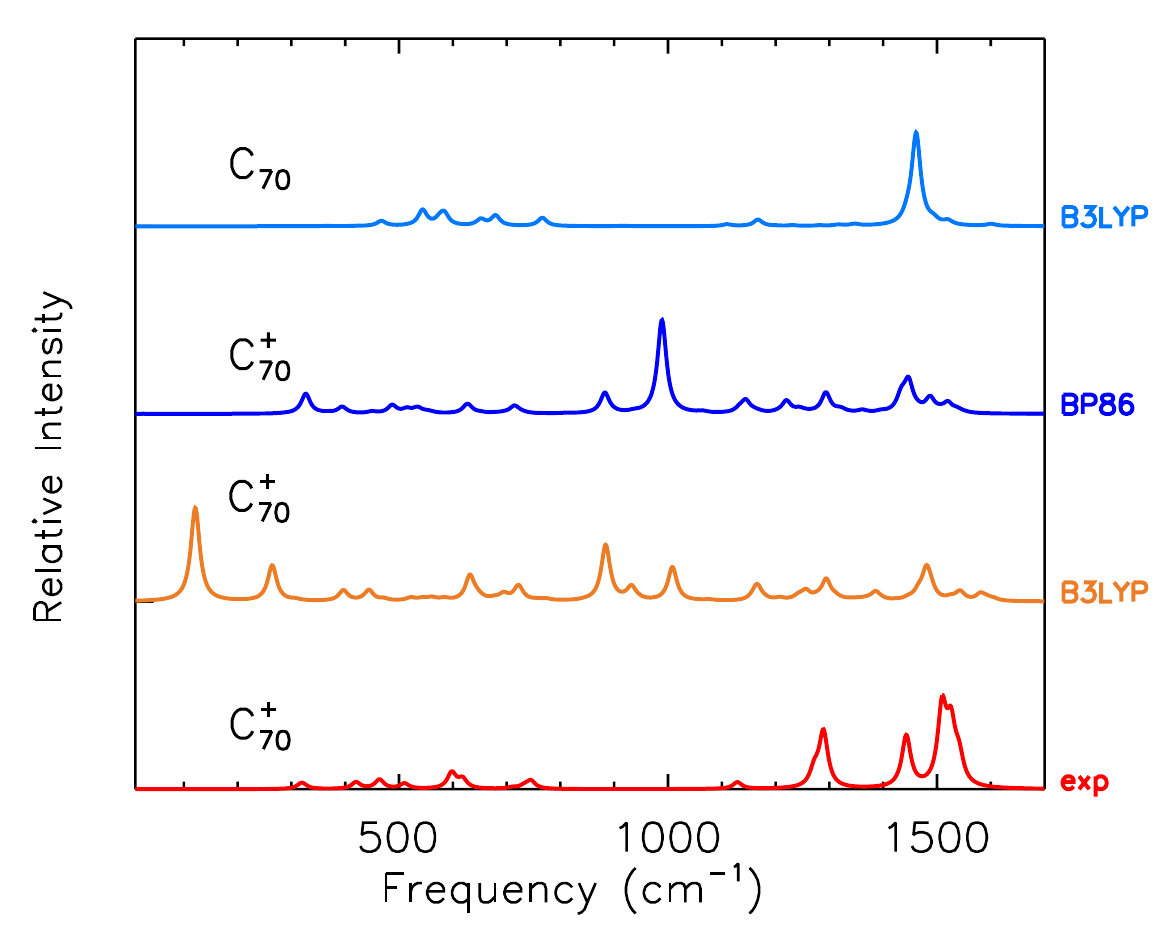}
\caption{Comparison between the experimental IR spectrum of C$_{70}^+$
   obtained by \citet{Kern2016} and theoretical IR spectra
  calculated with B3LYP/6-31G* and BP86/6-31G*  for the
  correct $C_{2v}$ structure. The B3LYP/6-31G* spectrum of neutral
  C$_{70}$ is also shown for comparison. The spectra are convolved
  with a Lorentzian function with a FWHM of 10 \wav and normalised to
  the strongest feature.}
\label{fig:C70+}
\end{figure*}

\citet{Bausch2010} documented the failure of the B3LYP functional for
open-shell cationic PAHs including 5-membered rings. These authors
suggested using the BP86 functional to verify the calculated
frequencies and intensities. We followed their suggestion and
calculated the BP86/6-31G* IR spectrum of
C$_{70}^+$. Fig.~\ref{fig:C70+} compares the experimental
\citep{Kern2016} and two calculated spectra of C$_{70}^+$ for the
correct structure of $C_{2v}$ symmetry. The BP86 spectrum has an
intense feature at 1000~\wav which is not observed in the experimental
spectrum. In general, positions of the other modes are in better
agreement with experimental values than for the B3LYP
spectrum. \citet{Kern2016} reached a similar conclusion even if the
symmetry of their structure was lower ($C_s$ and $C_i$ instead of
$C_{2v}$). \citet{Popov} calculated the vibrational frequencies of
C$_{70}^+$ with the PBE/TZ2P methods and their results are in better
agreement with our BP86/6-31G* frequencies than with the B3LYP/6-31G*
results. Interestingly, the strongest B3LYP/6-31G* peak at 121~\wav
corresponds to the splitting of the $E_2'$ mode of C$_{70}$ which
arises due to the Jahn--Teller distortion of C$_{70}^+$. Contributions
from other spin multiplicity states can be ruled out, since the
quadruplet lies 1.5~eV higher than the doublet.  A possible
explanation of these anomalous calculated spectra may lie in the
breakdown of the Born--Oppenheimer approximation due to the
presence of excited states at energies comparable to vibrational
energies. 
 It is thus clear that, for some fullerene ions such as  C$_{70}^+$, calculations of IR spectra based on the harmonic approximation to a single potential energy surface are unreliable and should be interpreted with great caution.

\subsection{Comparison with astronomical observations}

C$_n$ cages with $n=44$, 50, 56, 62, 64, and 68 are likely
contributors to the IR emission of the astrophysical environments
known to contain significant amounts of C$_{60}$. To assess the
contribution of those structures, we compared their IR spectra to the
observations of the three fullerene-rich PNe studied in
\citet{Jero:C60excitation}: the galactic PN Tc 1, and SMP~SMC~16
(SMC16 hereafter) and SMP~LMC~56 (LMC56 hereafter) -- similar objects
in the Magellanic Clouds. These observations show \csixty emission,
whereas our theoretical calculations pertain to absorption
spectra. The emission process for these species starts by absorption
of a UV photon; this is followed by a rapid iso-energetic transition
to the electronic ground state, but the process leaves the molecules
in a highly excited vibrational state. Fluorescent IR emission then
cools the molecule, and this is the emission that we observe. It is
possible to carry out a full calculation for this IR fluorescent
emission, but for our purposes we will use an approximation that is
simpler and faster. Indeed, the full fluorescence
spectrum can be approximated by multiplying the calculated intensities
by the Planck function at a well-chosen temperature; we used $T=750$~K
\citep[see e.g.][]{Bauschlicher:PAHdb}. A full numerical fluorescence
calculation would primarily differ in providing different relative
intensities between the shortest and the longest wavelength
modes. Given the uncertainties associated the intrinsic IR
intensities and with details of the emission mechanism, this
approximation is thus justified for our purposes. We also convolved
the resulting emission peaks with a Gaussian profile with a FWHM of
8~\wav,  the width of the observed C$_{60}$ bands
\citep{Cami:C60-Science}. Since the PNe mentioned above do not show
any evidence of C$_{60}^+$, we focus our comparison on the neutral
species.

\begin{figure}
\centering
\includegraphics[scale=0.7]{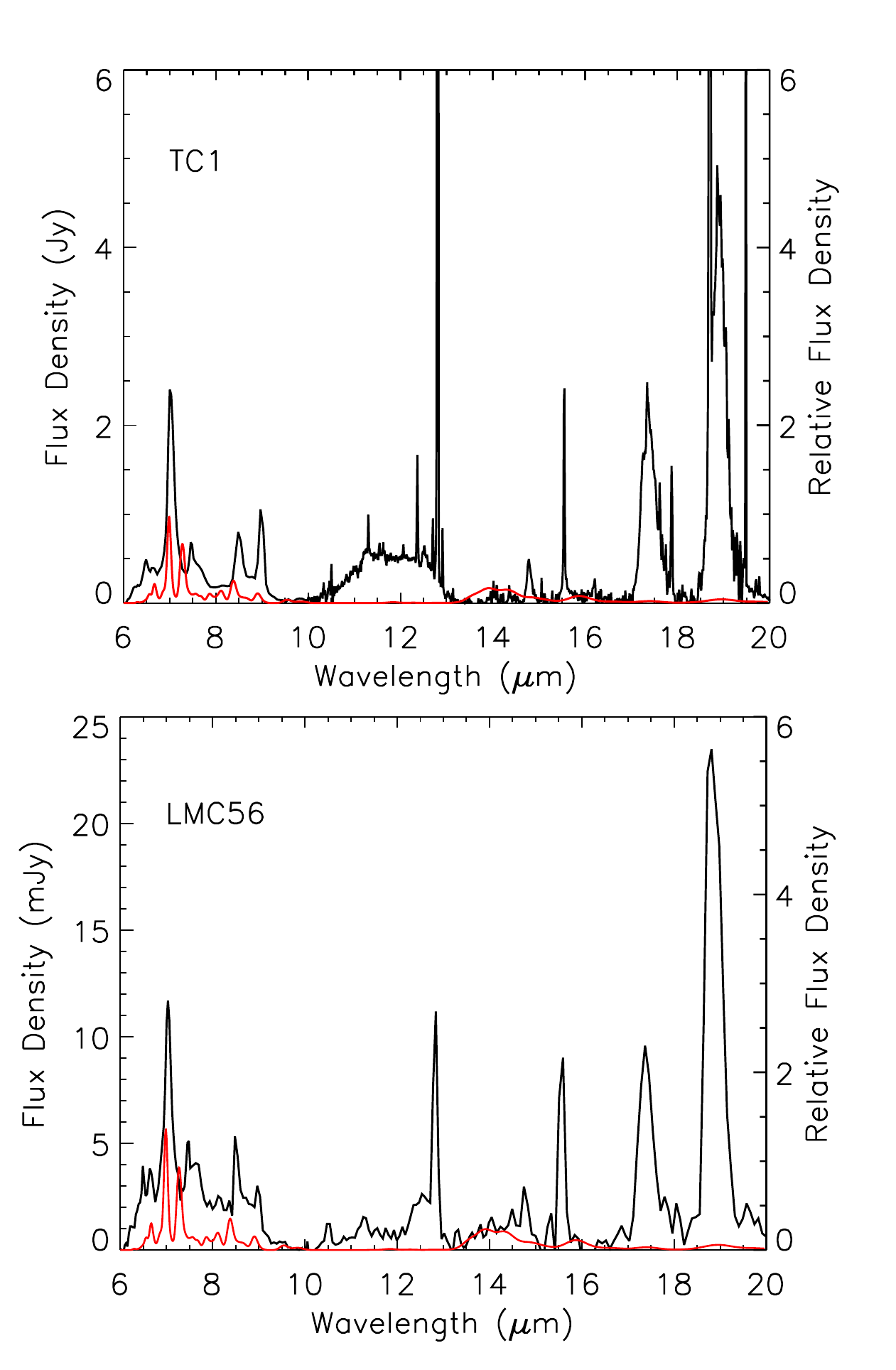}
\caption{Observed Spitzer/IRS mid-IR spectra (black) of Tc~1 and
  SMC~56 (from \citealt{Jero:C60excitation}) compared to the summed
  emission spectrum (red) of C$_{44}$ (isomers 72, 75, 89), C$_{50}$
  (isomers 263, 270, 271), and C$_{60}$.}
\label{Fig:smallcages_obs}
\end{figure}

\begin{figure}
\centering
\includegraphics[scale=0.7]{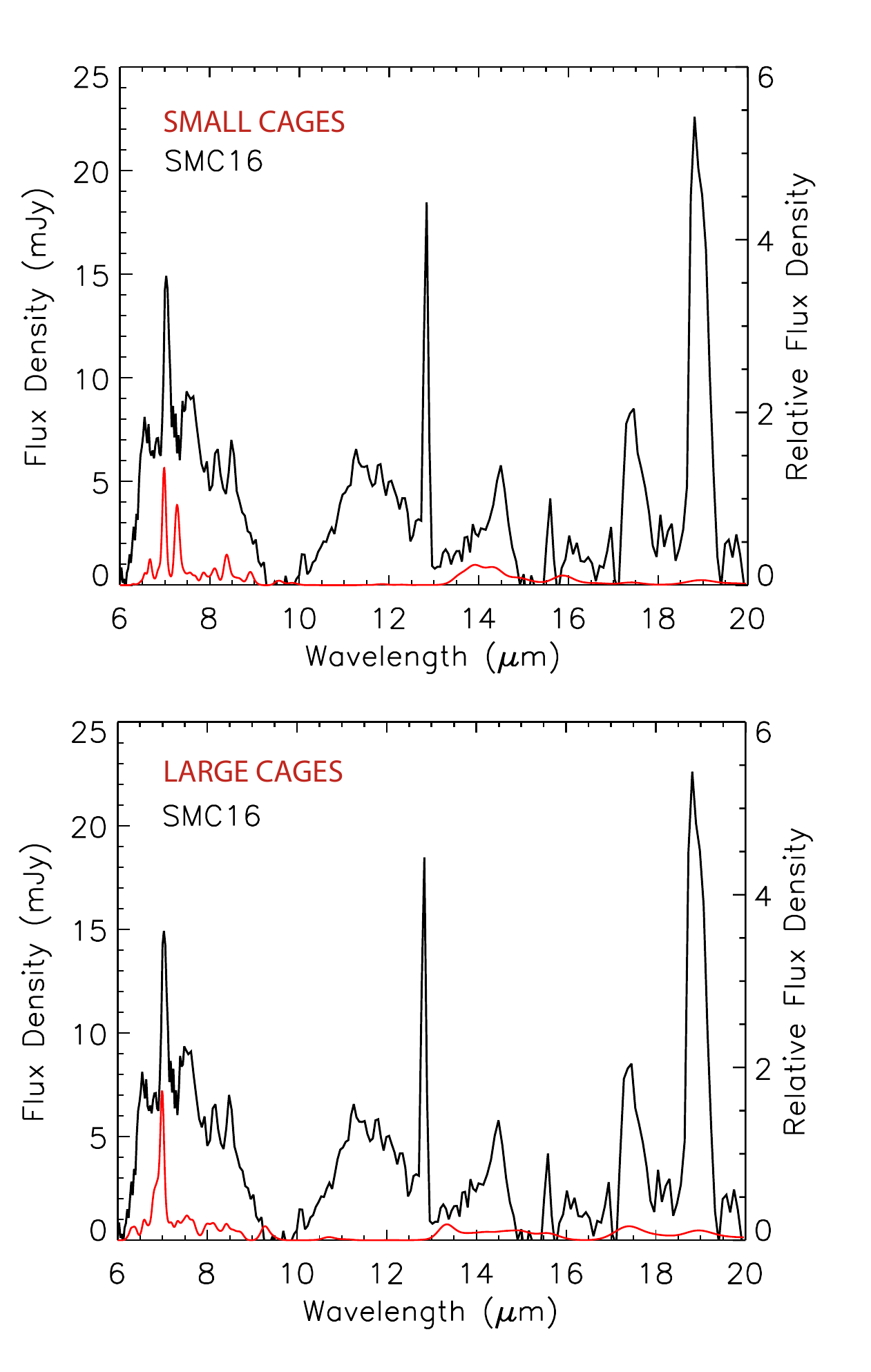}
\caption{Top panel: Observed Spitzer/IRS mid-IR spectrum of SMC~16
  (black) from \citet{Jero:C60excitation} compared to the summed
  emission spectrum (red) of C$_{44}$ (isomers 72, 75, 89), C$_{50}$
  (isomers 263, 270, 271), and C$_{60}$. Bottom panel: The same
  observational spectrum of SMC~16 compared to the summed emission
  spectrum of C$_{64}$ (isomers 3451, 3452), C$_{68}$ (isomers 6290,
  6328, 6270), C$_{60}$, and C$_{70}$.}
\label{Fig:largecages_obs}
\end{figure}

The observed IR spectra of SMC16 and LMC56 have several features in
the 770--665~cm$^{-1}$ (13--15$~\mu$m) region\footnote{We note that
  there is a well-known artifact of the Spitzer IR spectrograph
  between 13.2 and 14 $\mu$m, called the ``teardrop", which shows up
  in observations through the SL1 module; the features discussed here
   originate from the LL2 module and hence cannot be attributed
  to the teardrop artifact.}, where most of the C$_n$ cages studied in
this work show activity. Furthermore, the detailed structure of the
peaks in the 6--8~$\mu$m range of the observed spectrum is well
reproduced by the summed spectra of the small cages
(Fig.~\ref{Fig:smallcages_obs}). Nevertheless, the correlation between
the observed and simulated spectra is too tenuous to prove the
presence of any specific C$_n$ cage in these astronomical
objects. Such assignments are made especially challenging by the fact
that the abundance of non-C$_{60}$ cages is expected to be much lower
than that of C$_{60}$; in addition, intrinsic intensities of the
IR-active modes of other fullerenes are lower than those of
C$_{60}$. Thus, individual species should have very weak emission
features at best.

If we cannot detect individual species, could we perhaps detect a
spectral fingerprint of an entire cage population?
Figure~\ref{Fig:largecages_obs} shows simulated summed emission
spectra of small and large C$_n$ cages, in each case compared to the
observed IR spectrum of SMC 16. The smaller cages show features
in the 13--15~$\mu$m region and intriguingly, the average spectrum of
the smaller cages shows a weak, broad feature seen in both LMC~56 and SMC~16 without overestimating the flux at
any other wavelength (Note that for a lower temperature in our
emission model, the emission peaks in the 6--9 $\mu$m region would
be weaker). It is thus conceivable that a population of smaller cages is
indeed present in these objects. The observations are compatible
with such a population.

A similar average spectrum of the most stable larger cages (64 to 70
carbon atoms) also yields a small feature in the 13--15~$\mu$m region
that is weaker and closer to 13~$\mu$m. The summed spectra also shows
a mode around 7~$\mu$m (1728~cm$^{-1}$), which can be attributed to
C$_{50}$ (isomer 271) and C$_{70}$, and would overlap with the
corresponding \csixty band. Here too, the observations are compatible
with a population of larger cages, but they offer no robust detection.

Finally, it is worth pointing out that the cages studied here have
IR-active modes between 6-9~$\mu$m corresponding to a strong plateau
that is seen underneath the emission features in all three
astronomical objects. A population of smaller and/or larger cages
could thus also be partly responsible for this plateau emission. On
the other hand, little to no emission is seen in the 10-13~$\mu$m
region, where another plateau appears in the observations.  Therefore, this
emission plateau cannot be due to fullerenes. 

\section{Conclusions}

We have investigated the stability of fullerene cages with 44 to 70
carbon atoms and calculated their IR-active vibrational modes. The IR
spectra of different isomers of the same species are similar, and
therefore it is difficult to identify  specific isomers based on their
IR spectra, except perhaps  for the highly symmetric species such as
C$_{50}$ isomer 271. The spectra of most cages
that are smaller than C$_{60}$ all show features in the 13--15~$\mu$m
range, where the astronomical spectra of fullerene-rich planetary
nebulae also contain characteristic signals. We find that the
astronomical observations are compatible with the presence of a
population of fullerene cages, but offer no robust evidence for
them. Better chances to identify these species will be possible when
the James Webb Space Telescope, to be launched in 2021, will provide
high-quality spectra of C$_{60}$-containing objects. The theoretical
spectra presented here will be useful in interpreting those
data. Finally, we want to point out that the presence of low-lying
electronic states for some highly symmetric fullerene ions can greatly complicate
matters and that their vibrational spectra calculated in the Born-Oppenheimer and harmonic approximations should be treated with cautious distrust.

\section*{Acknowledgements}

AC acknowledge financial support from the Netherlands
Organisation for Scientific research (NWO) through a Veni grant
(No.~639.041.543). Calculations were carried out on the Dutch national
e-infrastructure (Cartesius) with the support of SURF Cooperative,
under NWO-EW project SH-362-15. Part of the research was performed under
the Leiden/ESA Astrophysics Program for Summer Students (LEAPS). JC,
EP, and VNS acknowledge support from NSERC through the Discovery Grants
Program. HM was supported by an interdisciplinary undergraduate research
award from the Centre for Planetary Science and Space Exploration (CPSX)
at the University of Western Ontario.


\newpage
\bibliographystyle{mnras.bst}
\bibliography{fullerenes} 

\bsp	
\label{lastpage}
\end{document}